\documentclass[letter]{aa} 

\usepackage{graphicx}
\usepackage{txfonts}
\usepackage{amstext}

\begin{document}

   \title{Radio detection of VIK~J2318$-$3113, the most distant radio-loud quasar ($z$=6.44)}


   \author{L. Ighina
          \inst{1,2}
          \and
          S. Belladitta\inst{1,2}
          \and
          A. Caccianiga\inst{1}
          \and
          J. W. Broderick \inst{3}
          \and
          G. Drouart\inst{3}
          \and
          A. Moretti\inst{1}
          \and
          N. Seymour\inst{3}
          }

   \institute{INAF - Osservatorio Astronomico di Brera, via Brera 28, 20121 Milan, Italy\\
              \email{lighina@uninsubria.it}
         \and
             DiSAT - Università degli Studi dell' Insubria, via Valleggio 11, 22100, Como, Italy
        \and
        International Centre for Radio Astronomy Research, Curtin University, 1 Turner Avenue, Bentley, WA 6102, Australia
             }

   \date{}

\abstract{We report the 888 MHz radio detection in the Rapid ASKAP Continuum Survey (RACS) of VIK~J2318$-$3113, a $z$=6.44 quasar. Its radio luminosity (1.2 $\times 10^{26}$ W Hz$^{-1}$ at 5~GHz) compared to the optical luminosity (1.8 $\times 10^{24}$ W Hz$^{-1}$ at 4400~\AA) makes it the most distant radio-loud quasar observed so far, with a radio loudness R$\sim$70 (R$=L_\mathrm{{5GHz}}/L_\mathrm{{4400\AA}}$). Moreover, the high bolometric luminosity of the source (L$_\mathrm{{bol}}$=7.4 $\times 10^{46}$ erg s$^{-1}$) suggests the presence of a supermassive black hole with a high mass ($\gtrsim$6 $\times$10$^8$ M$_\odot$) at a time when the Universe was younger than a billion years. Combining the new radio data from RACS with previous ASKAP observations at the same frequency, we found that the flux density of the  source may have varied by a factor of $\sim$2, which could suggest the presence of a relativistic jet oriented towards the line of sight, that is, a blazar nature.
However, currently available radio data do not allow us to firmly characterise the orientation of the source. Further radio and X-ray observations are needed.
} 


   \maketitle


\section{Introduction}
In recent years, the exploitation of numerous optical and infrared (IR) wide-area surveys (e.g. the Panoramic Survey Telescope and Rapid Response System, Pan-STARRS, \citealt{Chambers2016}; the VISTA Kilo-degree Infrared Galaxy Survey, VIKING, \citealt{Edge2013}; the Dark Energy Survey, DES, \citealt{Abbott2016}, etc.) has led to the discovery of thousands of high-$z$ quasars (QSOs), with more than 200 sources discovered at $z$\textgreater6 (e.g. \citealt{Mazzucchelli2017,Matsuoka2019a,Fan2019,Wang2019,Andika2020}), the three most distant of which are at \textit{z}$\sim$7.5 \citep{Banados2018,Yang2020,Wang2021}.
These sources have already proved to be very useful tools for investigating the intergalactic medium (IGM) at early cosmic times through the absorption of their optical spectra bluewards of Ly$\alpha$ (e.g. \citealt{Kashikawa2006,Gaikwad2020}).
Moreover, the mere presence of such powerful and massive objects in the primordial Universe places strong constraints on theoretical models describing the evolution and the accretion rate of supermassive black holes (SMBHs; e.g. \citealt{Volonteri2015,Wang2020}).\\
Decades of studies at low redshift have now established that radio-loud (RL\footnote{We considered a QSO to be radio loud when it has a radio loudness $R$\textgreater10, with $R$ defined as the ratio of the 5~GHz and 4400~\AA ~rest-frame flux densities, $R=S_{\mathrm{5 GHz}}/S_\mathrm{{4400 \AA}}$ \citep{Kellerman1989}.}) sources represent $\sim$10-15\% of the total QSO population (e.g. \citealt{Retana2017}), with no significant deviations until \textit{z}$\sim$6 (e.g. \citealt{Stern2000,Liu2021}; Diana et al. in prep.). However, of all the $z$\textgreater6 QSOs, only a few have a radio detection, which means that there are far fewer confirmed high-$z$ RL QSOs. 
To date, only five have been found at $z$\textgreater6 \citep{McGreer2006,Banados2015,Belladitta2020,Liu2021}, with the most distant being at $z$=6.21 (\citealt{Willot2010}). 
As described by \cite{Kellermann2016}, the RL classification (R\textgreater10), as opposed to the radio quiet (RQ; R\textless10), should identify sources that produce the radio emission through a relativistic jet, which can significantly affect both the accretion process itself and the environment of the source (see \citealt{Blandford2019} for a recent review).
Identifying and characterising powerful RL sources at the highest redshifts therefore is of key importance for studying the role of relativistic jets in the primordial Universe.\\
In this Letter we report the radio detection of the $z$=6.444$\pm$0.005 QSO VIKING~J231818.35$-$311346.3 (hereafter VIK~J2318$-$3113; \citealt{Decarli2018}). With a relatively bright radio flux density ($\sim$1.4~mJy at 888~MHz), this source is the most distant RL QSO observed  to date. 
VIK~J2318$-$3113 was discovered from the near-IR (NIR) VIKING survey with the dropout technique, and its redshift was confirmed with both X-Shooter in the NIR and the Atacama Large Millimetre/submillimetre Array (ALMA) in the submillimetre \citep{Decarli2018, Yang2020b}. In this Letter we present its radio properties using recent observations, and by combining them with archival data, we also compare VIK~J2318$-$3113 with the small number of other high-$z$ RL QSOs.\\
We use a flat $\Lambda$CDM cosmology with $H_{0}$=70 km s$^{-1}$ Mpc$^{-1}$, $\Omega_m$=0.3, and $\Omega_{\Lambda}$=0.7. Spectral indices are given assuming $S_{\nu}\propto \nu^{-\alpha}$ , and all errors are reported at 1$\sigma$ unless otherwise specified.\\

\section{Radio observations}

\subsection{888~MHz ASKAP observations}
\label{subsec:888}
VIK~J2318$-$3113 has been detected in the first data release of the Rapid ASKAP Continuum Survey (RACS; \citealt{McConnell2020})\footnote{\url{https://data.csiro.au/collections/collection/CIcsiro:46532.}} with a peak flux density of 1.43 mJy beam$^{-1}$at 888~MHz, which considering the associated RMS  (0.19 mJy beam$^{-1}$), corresponds to a signal-to-noise ratio (S/N) \textgreater7 (values as reported in the catalogue released on 2020 December 17).\\
The overall RACS survey is planned to cover the entire sky south of declination $+51^{\circ}$ (36656~deg$^{2}$ in total) in three different radio bands centred at 888, 1296, and 1656~MHz, all with a bandwidth of 288~MHz. 
These observations are designed as a pilot project to prepare for the data calibration and handling of future deeper surveys (e.g. the evolutionary map of the Universe, EMU, \citealt{Norris2011}) with the Australian SKA Pathfinder (ASKAP; \citealt{Johnson2008}).
In the first data release (December 2020), the sky south of declination $+41^{\circ}$ was covered in the lower frequency band (888~MHz) with a spatial resolution of $\sim$15$''$.
By cross-matching this first data release with the list of  $z$\textgreater6 QSOs discovered to date in the same sky area (169 sources in total), we found the radio counterparts of three of them: VIK~J2318$-$3113, and two other RL QSOs. For these last two objects a discussion of their radio properties has already been reported in the literature: FIRST~J1427385+331241 ($z$=6.12; \citealt{McGreer2006}) and PSO~J030947.49+271757.31 ($z$=6.10; \citealt{Belladitta2020}). 
\begin{figure}
   \centering
   \includegraphics[width=\hsize]{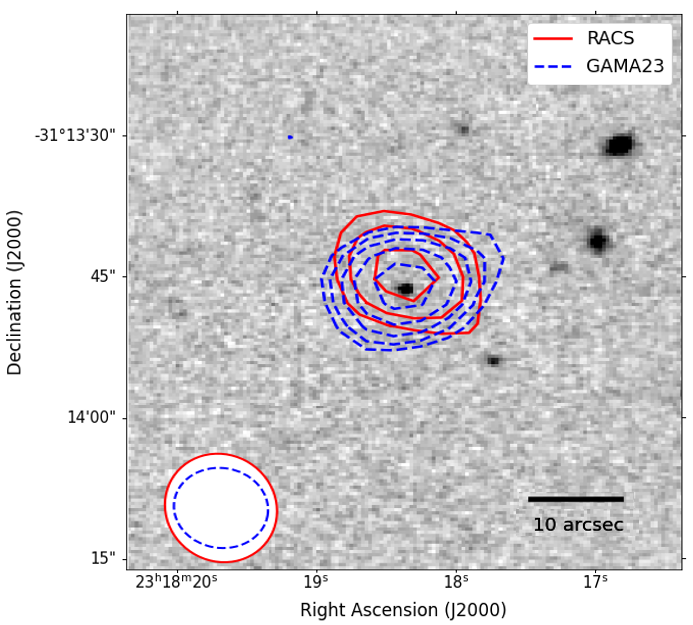}
      \caption{1$'$ $\times$ 1$'$ cutout of the $Y$-band VIKING image around VIK~J2318$-$3113, overlaid with the 888~MHz radio contours from RACS (continuous red lines) and GAMA23 (dashed blue lines). In both cases the contours are spaced by $\sqrt{2}$ starting from $\text{three times}$ the off-source RMS derived in our analysis, $\sim$0.20~mJy~beam$^{-1}$ for RACS and $\sim$0.04~mJy~beam$^{-1}$ for GAMA23. In the bottom left corner the beam sizes from the RACS (12.2$''$ $\times$ 11.4$''$) and GAMA23 (10.2$''$ $\times$ 8.5$''$) observations are shown.}
         \label{fig:opt_radio}
\end{figure}
The radio source is located 1.6$''$ from the optical/NIR counterpart of VIK~J2318$-$3113, which is consistent with the positional error reported in the RACS catalogue ($\sim$4$''$).
Even considering typical uncertainties in interferometric radio positions ($\approx$$\frac{\Delta \theta}{2\times S/N}$$\sim$0.9$''$, where $\Delta \theta$ is the size of the beam; \citealt{Fomalont1999}) together with the typical astrometric precision of the survey ($\sim$0.8$''$; \citealt{McConnell2020}), the observed offset is still consistent. Moreover, from the source density of the RACS survey ($\sim$80 sources deg$^{-2}$, \citealt{McConnell2020}), we can also compute the probability of finding an unrelated radio source within a 1.6$''$ radius from any given position (see e.g. eq. 4 in \citealt{Condon1998}). In this case, the probability is $\sim$5$\times$10$^{-5}$, which means that the expected number of spurious associations of the 169 $z$\textgreater6 QSOs  that we based the query on is \textless0.01.
We can therefore conclude that the association between VIK~J2318$-$3113 and the radio source is statistically significant and unlikely to be spurious.\\
At the same time, VIK~J2318$-$3113 also belongs to one of the Galaxy and Mass Assembly (GAMA; \citealt{Driver2011}) fields, GAMA23 (339 \textless\, R.A. [deg]  \textless\, 351 and $-$35 \textless \, Dec. [deg] \textless\,$-$30). In particular, this region has recently (2019 March) been covered by a deeper ASKAP observation (RMS$\sim$0.04 mJy beam$^{-1}$), again at 888~MHz, within an ASKAP/EMU early science project\footnote{\url{https://data.csiro.au/collections/collection/CIcsiro:40262.}} and was reduced as described in \cite{Seymour2020}. We report in Fig. \ref{fig:opt_radio} the 888~MHz radio contours from the RACS and GAMA23 observations, overlaid on the NIR VIKING image in the $Y$-band.
\begin{table}
\centering
\caption{Results of the analysis of the 888~MHz ASKAP observations of VIK~J2318$-$3113.}

\begin{tabular}{lcccccc}
\hline
\hline
    Project: & RACS & GAMA23\\
    Total flux density (mJy): & 1.44$\pm$0.34$^{*}$ & 0.59$\pm$0.07\\
    Peak surf.  brightness (mJy/beam):  & 1.48$\pm$0.20 & 0.59$\pm$0.04\\
    Major axis$^{**}$ (arcsec): & 13.2$\pm$2.0 & 10.5$\pm$0.8\\
    Minor axis$^{**}$ (arcsec): & 10.2$\pm$1.2 & 8.2$\pm$0.5 \\
    P.A. east of north (deg): & 45$\pm$18 & 105$\pm$10\\
    Off-source RMS (mJy/beam): & 0.20 & 0.04\\
    
\hline
\hline
\end{tabular}
\label{tab:j2318gaussfit}

\tablefoot{
\tablefoottext{*}{In the following we use the more conservative error of 0.60~mJy obtained from eq. 7 in \cite{McConnell2020}. See section \ref{subsec:888} for further details.}\\
\tablefoottext{**}{Convolved with the beam of the instrument.}

}

\end{table}
In Tab. \ref{tab:j2318gaussfit} we report the results of a single Gaussian fit performed on the RACS and GAMA23 images using the Common Astronomy Software Applications package (CASA; \citealt{Mcmullin2007}). In the GAMA23 observation the best-fit position is only 0.37$''$ away from the NIR counterpart, thus providing further strong evidence for the radio association. Given the very similar angular resolution in both cases, the source is point-like and not resolved. However, the estimated flux density varies by a factor $\sim$2.4 in the two images, from 0.59$\pm$0.07 to 1.44$\pm$0.34 mJy. \\
The time separation between the two observations is one year (2019 March -- 2020 March), which in the source rest frame corresponds to $\sim$50 days (without taking possible relativistic effects into account).
In order to verify whether the source variation between the GAMA23 and RACS observations is real or is only a systematic effect related to the calibration, we compared the integrated flux densities of the sources detected in the two images. In particular, as for VIK~J2318$-$3113, we performed a single Gaussian fit with the CASA software on $\sim$70 sources with a flux density between 1 and 10~mJy and within one square degree from the QSO position\footnote{Although a primary beam correction was performed during the data reduction of the RACS survey and the GAMA23 images, we applied a search radius cutoff in order avoid any possible residual fluctuation of the flux calibration.}.
The distribution of the ratios of the flux densities measured in the two images is a Gaussian centred at one and with $\sigma$=0.16, consistent with the statistical errors on the flux densities and thus indicating that the observed difference for VIK~J2318$-$3113 cannot be attributed to a systematic calibration offset in the two datasets. When we sum in quadrature the uncertainties related to the two flux density estimates, the significance of the variation observed in VIK~J2318$-$3113 is $\sim$2.4$\sigma$. A large variation in a short period of time as observed in this case is usually associated with the presence of a relativistic jet oriented towards the line of sight, that is, a blazar nature (e.g. \citealt{Hovatta2008}).\\
The uncertainty on the flux density ratios reported above ($\sigma$=0.16) was derived from the relative comparison of the RACS and GAMA23 images, that is, from datasets obtained from the same telescope. \cite{McConnell2020} have studied the uncertainties on the absolute flux density scale of RACS images by comparing sources with multiple independent RACS observations (i.e. on the overlapping edges of different tiles), also with other catalogues in the literature, finding $\Delta$S$_\nu$ = 0.5~mJy + 0.07$\times$S$_\nu$ (eq. 7 in their paper). In the particular case of VIK J2318$-$3113, the corresponding value is $\sim$0.60 mJy. We take this uncertainty into account in section \ref{sec:radio_loud} when we compute the quantities based on the RACS flux density (e.g. radio luminosity and radio loudness).

\subsection{Archival radio observations}
Even though VIK~J2318$-$3113 is not reported in any other public radio catalogue, we checked archival radio images at the NIR position of the source to search for the presence of a faint but significant (S/N\textgreater2.5) radio signal.
We did not detect the source in the TIFR Giant Metrewave Radio Telescope Sky Survey (TGSS; \citealt{Intema2017}) at 150~MHz (image RMS$\sim$2.9~mJy beam$^{-1}$), the Sydney University Molonglo Sky Survey (SUMSS; \citealt{Mauch2003}) at 843~MHz (image RMS$\sim$2.5~mJy beam$^{-1}$), or in the NRAO Karl G. Jansky Very Large Array Sky Survey (NVSS; \citealt{Condon1998}) at 1.4~GHz (image RMS$\sim$0.45~mJy beam$^{-1}$).
In contrast, we did find a radio excess less than 0.6$''$ away from the NIR position of the source in the first (2018 February) and second (2020 November) epochs of the Very Large Array Sky Survey (VLASS; \citealt{Lacy2020}) at 3~GHz. The peak flux density of the emission in the two epochs is 0.29$\pm$0.11~mJy~beam$^{-1}$ in the first and 0.40$\pm$0.13~mJy~beam$^{-1}$ in the second, which corresponds to a S/N of 2.6 and 3.0, respectively. Even though the two estimates are marginally consistent, we consider the average of the two and the overall range of uncertainty because of the possible intrinsic variability of the source: 0.35$\pm$0.18~mJy.
In Tab. \ref{tab:radio_data} we report the radio data and the 2.5$\sigma$ upper limits obtained from archival observations as described above.\\
When we take currently available data with their uncertainties and the upper limits derived from non-detections into account, the spectral index of a single power law covering the observed frequency range is poorly constrained ($\alpha_r$=0--1.2).
However, in addition to information on the flux density and the dimensions of the sources, the RACS catalogue also reports the spectral index computed within the 288~MHz band centred at 888~MHz. The spectral index reported for VIK~J2318$-$3113 is $\alpha_r$=0.98, which is similar to what is typically observed in high-$z$ QSOs (e.g. \citealt{Coppejans2017,Banados2018}). In the following, we consider this to be the best-fit value despite the relatively low S/N across the ASKAP band, even though a different assumption does not affect the results.
A more detailed discussion of the broad-band radio properties of VIK~J2318$-$3113 will be presented in a forthcoming work.

\begin{table}
\centering
\caption{Estimates and 2.5$\sigma$ upper limits on the radio flux densities of VIK~J2318$-$3113 from archival radio surveys.}
\label{tab:radio_data}
\setlength{\tabcolsep}{3.pt}
\begin{tabular}{lcccccccc}
\hline
\hline
        Survey: & TGSS  & SUMSS & NVSS & VLASS \\
                Obs. Freq. (GHz): & 0.15 & 0.843 &1.4 & 3 \\
                Flux density (mJy/beam): & \textless7.3 & \textless 6.3 & \textless1.1& 0.35$\pm$0.18\\
                
\hline
\hline
\end{tabular}
\end{table}


\section{Optical/UV properties}
Given the high-redshift nature of VIK~J2318$-$3113, the NIR photometric data from the VIKING survey (reported in Tab. \ref{tab:nir}) cover the UV/optical spectrum in its rest frame. Therefore we used these photometric points to  estimate the bolometric luminosity (L$_\mathrm{{bol}}$) of the source. In the following, we assume an optical spectral index given by the slope observed between the $K$ and $J$ bands, $\alpha_{ {\mathrm{{o}}}}=0.54$,  which is consistent with what is normally found in other QSOs (e.g. \citealt{Vandenberk2001}). 
We started by computing the rest-frame monochromatic luminosities at 1350 and 3000~\AA~ using the observed magnitudes in the filter with the closest corresponding rest-frame wavelength, that is, $Y$ ($\sim$1370~\AA) and $K$ ($\sim$2860~\AA), respectively. The bolometric luminosity can then be inferred using the correction factors derived in \cite{Shen2008} for 1350~\AA \, ($L_\mathrm{{bol}}$= 8.0 $\pm$ 2.7 $\times$ 10$^{46}$~erg~s$^{-1}$) and in \cite{Runnoe2012} for 3000~\AA \, ($L_\mathrm{{bol}}$= 7.4 $\pm$ 0.3 $\times$ 10$^{46}$~erg~s$^{-1}$). 
Averaging the two results with the corresponding variances as weights, we obtain $L_\mathrm{{bol}}$= 7.4 $\pm$ 0.3 $\times$ 10$^{46}$~erg~s$^{-1}$.
Assuming an Eddington-limited accretion, that is, L$_\mathrm{{bol}}$ $\leq$L$_\mathrm{{EDD}}$,\footnote{Where L$_\mathrm{{EDD}}$ = 1.26 $\times$ 10$^{38}$ (M$_\mathrm{{BH}}$/M$_{\odot}$)~erg~s$^{-1}$.} this value of the bolometric luminosity implies that the SMBH mass must be higher than 6 $\times~10^{8}$ M$_{\odot}$.

\begin{table}[h]
\centering
\caption{NIR magnitudes of VIK~J2318$-$3113 as measured in the VIKING survey (Vega system).}
\label{tab:nir}
\begin{tabular}{lccccc}
\hline
\hline
        Filter: & $Z$ & $Y$ & $J$ & $H$ & $K$\\
                $\lambda_{eff}$ ($\mu$m): &0.878 &      1.021 & 1.254 & 1.646 & 2.149 \\
                magnitude: & 21.42 & 20.17 & 19.89 & 19.61 & 18.67\\
                mag. error: & 0.11 & 0.08 & 0.11 & 0.18 & 0.14 \\
\hline
\hline
\end{tabular}
\end{table}

\section{Radio loudness and comparison with high-$z$ RL QSOs}
\label{sec:radio_loud}
In order to estimate the rest-frame monochromatic luminosity at 5~GHz, we considered the 888~MHz flux density obtained from the RACS observation (S$_{\mathrm{{888MHz}}}$= 1.44~mJy) with an uncertainty that takes the absolute calibration of the map into account (0.60~mJy, see previous section) and the spectral index reported in the RACS catalogue ($\alpha_r$=0.98). We also considered the GAMA23 flux density (S$_{\mathrm{{888MHz}}}$= 0.59$\pm$0.07~mJy) and a spectral index in the range $\alpha_r$=0--1.2 to estimate the associated uncertainty.
The result, however, has little dependence on the $\alpha_r$ assumption because the observed frequency of 888~MHz corresponds to a rest-frame frequency of 6.6~GHz, which is very close to 5~GHz. The resulting radio luminosity is L$_{\mathrm{{5GHz}}}$= 1.2$_{-0.9}^{+0.6} \, \times 10^{26}$ W~Hz$^{-1}$. Combining this estimate with the optical luminosity at 4400~\AA \, (L$_{\mathrm{{4400\AA}}}$= 1.8$\pm$0.1 $\times~10^{31}$ erg s$^{-1}$ Hz$^{-1}$), computed from the observed $K$ magnitude, we obtain a radio loudness R= 66.3$_{-46.7}^{+36.3}$. Adopting the typical value of R=10 as the threshold between RL and RQ sources, this makes VIK~J2318$-$3113 the most distant RL QSO observed so far, at $z$=6.44.
We note that this classification does not depend on the somewhat arbitrary criterion for separating the RL and RQ populations. Even if we consider a radio loudness as defined by \cite{Jiang2007}\footnote{In this case, the radio loudness is defined as the following rest-frame ratio: R=$S_{\mathrm{{5GHz}}}/S_{\mathrm{{2500\AA}}}$.} or a single threshold in the radio luminosity (L$_{\mathrm{{5GHz}}}$\textgreater $10^{32.5}$ erg s$^{-1}$ Hz$^{-1}$; \citealt{Jiang2007}), the RL classification still holds.\\
In RL QSOs, the radio emission is thought to be produced by relativistic jets and not by star-formation (SF)  processes (e.g. \citealt{Kellermann2016}). VIK~J2318$-$3113 was found to be very luminous in the far-IR (FIR) (log(L$_\mathrm{FIR}$/L$_\odot$) in the range 11.89--12.46, between 42.5 and 122.5 $\mu$m; \citealt{Decarli2018, Venemans2018, Venemans2020}), and this may imply that at least part of the observed radio emission is due to SF. However, considering the relation between radio and FIR luminosity observed in SF galaxies \citep{Condon2002}, we expect that only a few percent (\textless5\%) of the observed radio emission can be produced by SF. This confirms that the high radio power observed in VIK~J2318$-$3113 is likely produced by relativistic jets, as expected in RL sources. Interestingly, \cite{Venemans2020} also found that the FIR continuum and [C~II] emissions extend up to $\sim$5~kpc (0.2$''$) with an irregular morphology. Further radio observations at similar resolution would be fundamental for understanding the role of the different components at work in this complex QSO.\\
\begin{figure*}
   \centering
   \includegraphics[width=\hsize]{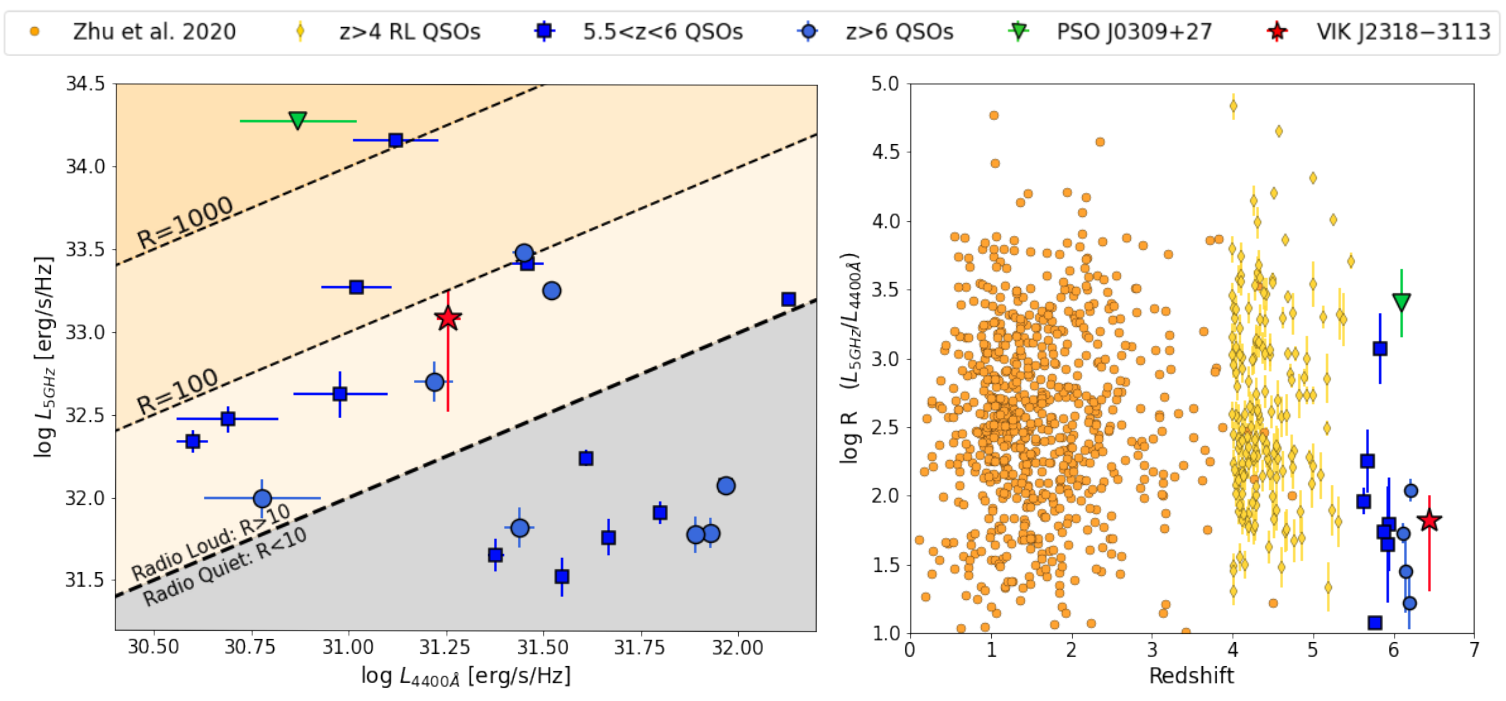}
      \caption{\textbf{Left:} Rest-frame radio luminosity density at 5~GHz vs. the rest-frame optical luminosity density at 4400~\AA~for $z$\textgreater5.5 QSOs with a radio detection in the literature. Diagonal lines indicate constant radio-loudness values. Adapted from \cite{Banados2015}. \textbf{Right:} Radio loudness as a function of redshift for the $z$\textgreater5.5 confirmed RL QSOs compared to an optically selected sample of RL QSOs at lower redshift (orange points; \citealt{Zhu2020}) and all the RL QSOs at $z$\textgreater4 (yellow diamonds) known to date. The blue squares (circles) report $z$\textgreater5.5 (\textgreater6) sources in both graphs. The only confirmed $z$\textgreater5.5 blazar \citep{Belladitta2020} is reported with a green triangle. At lower redshifts we did not distinguish this class because not all sources have a reliable classification. The red star represents VIK~J2318$-$3113.}
         \label{fig:R_diagram}
   \end{figure*}
Following \cite{Banados2015}, we report in Fig. \ref{fig:R_diagram} (left) the rest-frame radio luminosity (5~GHz) as a function of the rest-frame optical luminosity (4400~\AA) for the updated list of $z$\textgreater5.5 QSOs with a radio observation and thus a firm RL/RQ classification\footnote{Data from \cite{Banados2015,Banados2018}, \cite{Belladitta2020}, and \cite{Liu2021}}. Clearly, the radio loudness of VIK~J2318$-$3113 is similar to that of the majority of $z$\textgreater5.5 RL QSOs, with 10\textless R\textless100.\\
Moreover, in Fig. \ref{fig:R_diagram} (right) we compare the confirmed RL QSOs at $z$\textgreater5.5 to the optically selected sample at lower redshift ($\sim$800 sources) discussed in \cite{Zhu2020} and to the $z$\textgreater4 RL QSOs discovered so far\footnote{These are all the RL QSOs published to date. To estimate their radio loudness, we considered the radio spectral index, if present; otherwise, we assumed $\alpha_r$=0.75 \citep{Banados2015} and considered a $\pm$0.25 variation to estimate the uncertainty. The full list of sources with the corresponding radio data and references will be presented in Belladitta et al. (in prep.).}.
Interestingly,  only a small fraction of very radio-powerful
high-$z$ sources (logR\textgreater2.5) has been found at $z$\textgreater5.5  compared to low redshifts. This may be a consequence of the fact that at these redshifts, QSOs have been selected mainly in the optical/UV, with only three radio-selected sources (which include the two radio-brightest sources at $z$\textgreater6). Nevertheless, we expect that upcoming and ongoing wide-area surveys such as RACS and the development of dedicated selection techniques in the radio band (e.g. \citealt{Drouart2020})  will find many more radio-powerful sources at $z$\textgreater 6 (e.g. \citealt{Amarantidis2019}).


\section{Conclusions}
We have presented the radio detection (at 888 MHz) of VIK~J2318$-$3113, a $z$=6.44 QSO. Combining the new radio information from RACS with the archival data, we estimate a radio-loudness value of R$\sim$70, which means that this source is the most distant RL QSO observed to date. The radio association was made by cross-matching the first data release of the RACS survey and a list of the 169 previously discovered $z$\textgreater6 QSOs in the same area of the sky. As a result, we found radio counterparts for a total of three RL sources, VIK~J2318$-$3113 included, which corresponds to a radio detection rate of $\sim$2\% in the aforementioned list of $z$\textgreater6 QSOs. Because the RACS flux density limit is not deep enough to detect all the $z$\textgreater6 RL QSOs discovered so far, which have typical NIR magnitudes $\sim$22, this detection rate should be considered as a lower limit to the actual RL fraction at $z$\textgreater6.\\
We cannot fully characterise the radio spectral properties of VIK~J2318$-$3113, and thus establish whether it is a flat, steep, or peaked source, with the currently available radio data. This is an important diagnostic for understanding the orientation of the relativistic jet with respect to the line of sight, that is, whether VIK~J2318$-$3113 is a blazar. The possible presence of variability at 888~MHz, as found in the comparison of the RACS and GAMA23 observations, may suggest that the emission of this source is dominated by the relativistic beaming, which could mean that the jet is oriented at small angles from the line of sight. More data are required to confirm this result, however. 
Assuming an Eddington-limited accretion, the relatively high bolometric luminosity suggests the presence of a central SMBH with a mass $\gtrsim$6 $\times$ 10$^8 $M$_\odot$.\\
This detection anticipates the discovery of many more RL high-$z$ sources in the next years when the new generation of all-sky radio surveys will be performed by the Square Kilometre Array and its precursors.

\begin{acknowledgements}
We thank the anonymous referee for the useful comments and suggestions. We acknowledge financial contribution from the agreement ASI-INAF n. I/037/12/0 and n.2017-14-H.0 and from INAF under PRIN SKA/CTA FORECaST. In this work we have used data from the ASKAP observatory. The Australian SKA Pathfinder is part of the Australia Telescope National Facility which is managed by CSIRO. Operation of ASKAP is funded by the Australian Government with support from the National Collaborative Research Infrastructure Strategy. ASKAP uses the resources of the Pawsey Supercomputing Centre. Establishment of ASKAP, the Murchison Radio-astronomy Observatory and the Pawsey Supercomputing Centre are initiatives of the Australian Government, with support from the Government of Western Australia and the Science and Industry Endowment Fund. We acknowledge the Wajarri Yamatji people as the traditional owners of the Observatory site. This paper includes archived data obtained through the CSIRO ASKAP Science Data Archive, CASDA (\url{http://data.csiro.au}).
This research made use of Astropy (\url{http://www.astropy.org}) a community-developed core Python package for Astronomy \citep{astropy2013, astropy2018}. This research has made use of the SIMBAD database, operated at CDS, Strasbourg, France \citep{Simbad2000}.

\end{acknowledgements}

%
   \bibliographystyle{aa} 
   \bibliography{biblio} 

\begin{thebibliography}{53}
\expandafter\ifx\csname natexlab\endcsname\relax\def\natexlab#1{#1}\fi

\bibitem[{{Amarantidis} {et~al.}(2019){Amarantidis}, {Afonso}, {Messias},
  {Henriques}, {Griffin}, {Lacey}, {Lagos}, {Gonzalez-Perez}, {Dubois},
  {Volonteri}, {Matute}, {Pappalardo}, {Qin}, {Chary}, \&
  {Norris}}]{Amarantidis2019}
{Amarantidis}, S., {Afonso}, J., {Messias}, H., {et~al.} 2019, \mnras, 485,
  2694

\bibitem[{{Andika} {et~al.}(2020){Andika}, {Jahnke}, {Onoue}, {Ba{\~n}ados},
  {Mazzucchelli}, {Novak}, {Eilers}, {Venemans}, {Schindler}, {Walter},
  {Neeleman}, {Simcoe}, {Decarli}, {Farina}, {Marian}, {Pensabene}, {Cooper},
  \& {Rojas}}]{Andika2020}
{Andika}, I.~T., {Jahnke}, K., {Onoue}, M., {et~al.} 2020, \apj, 903, 34

\bibitem[{{Astropy Collaboration} {et~al.}(2018){Astropy Collaboration},
  {Price-Whelan}, {Sip{\H{o}}cz}, {G{\"u}nther}, {Lim}, {Crawford}, {Conseil},
  {Shupe}, {Craig}, {Dencheva}, {Ginsburg}, {VanderPlas}, {Bradley},
  {P{\'e}rez-Su{\'a}rez}, {de Val-Borro}, {Aldcroft}, {Cruz}, {Robitaille},
  {Tollerud}, {Ardelean}, {Babej}, {Bach}, {Bachetti}, {Bakanov}, {Bamford},
  {Barentsen}, {Barmby}, {Baumbach}, {Berry}, {Biscani}, {Boquien}, {Bostroem},
  {Bouma}, {Brammer}, {Bray}, {Breytenbach}, {Buddelmeijer}, {Burke},
  {Calderone}, {Cano Rodr{\'\i}guez}, {Cara}, {Cardoso}, {Cheedella}, {Copin},
  {Corrales}, {Crichton}, {D'Avella}, {Deil}, {Depagne}, {Dietrich}, {Donath},
  {Droettboom}, {Earl}, {Erben}, {Fabbro}, {Ferreira}, {Finethy}, {Fox},
  {Garrison}, {Gibbons}, {Goldstein}, {Gommers}, {Greco}, {Greenfield},
  {Groener}, {Grollier}, {Hagen}, {Hirst}, {Homeier}, {Horton}, {Hosseinzadeh},
  {Hu}, {Hunkeler}, {Ivezi{\'c}}, {Jain}, {Jenness}, {Kanarek}, {Kendrew},
  {Kern}, {Kerzendorf}, {Khvalko}, {King}, {Kirkby}, {Kulkarni}, {Kumar},
  {Lee}, {Lenz}, {Littlefair}, {Ma}, {Macleod}, {Mastropietro}, {McCully},
  {Montagnac}, {Morris}, {Mueller}, {Mumford}, {Muna}, {Murphy}, {Nelson},
  {Nguyen}, {Ninan}, {N{\"o}the}, {Ogaz}, {Oh}, {Parejko}, {Parley}, {Pascual},
  {Patil}, {Patil}, {Plunkett}, {Prochaska}, {Rastogi}, {Reddy Janga},
  {Sabater}, {Sakurikar}, {Seifert}, {Sherbert}, {Sherwood-Taylor}, {Shih},
  {Sick}, {Silbiger}, {Singanamalla}, {Singer}, {Sladen}, {Sooley},
  {Sornarajah}, {Streicher}, {Teuben}, {Thomas}, {Tremblay}, {Turner},
  {Terr{\'o}n}, {van Kerkwijk}, {de la Vega}, {Watkins}, {Weaver}, {Whitmore},
  {Woillez}, {Zabalza}, \& {Astropy Contributors}}]{astropy2018}
{Astropy Collaboration}, {Price-Whelan}, A.~M., {Sip{\H{o}}cz}, B.~M., {et~al.}
  2018, \aj, 156, 123

\bibitem[{{Astropy Collaboration} {et~al.}(2013){Astropy Collaboration},
  {Robitaille}, {Tollerud}, {Greenfield}, {Droettboom}, {Bray}, {Aldcroft},
  {Davis}, {Ginsburg}, {Price-Whelan}, {Kerzendorf}, {Conley}, {Crighton},
  {Barbary}, {Muna}, {Ferguson}, {Grollier}, {Parikh}, {Nair}, {Unther},
  {Deil}, {Woillez}, {Conseil}, {Kramer}, {Turner}, {Singer}, {Fox}, {Weaver},
  {Zabalza}, {Edwards}, {Azalee Bostroem}, {Burke}, {Casey}, {Crawford},
  {Dencheva}, {Ely}, {Jenness}, {Labrie}, {Lim}, {Pierfederici}, {Pontzen},
  {Ptak}, {Refsdal}, {Servillat}, \& {Streicher}}]{astropy2013}
{Astropy Collaboration}, {Robitaille}, T.~P., {Tollerud}, E.~J., {et~al.} 2013,
  \aap, 558, A33

\bibitem[{{Ba{\~n}ados} {et~al.}(2018){Ba{\~n}ados}, {Venemans},
  {Mazzucchelli}, {Farina}, {Walter}, {Wang}, {Decarli}, {Stern}, {Fan},
  {Davies}, {Hennawi}, {Simcoe}, {Turner}, {Rix}, {Yang}, {Kelson}, {Rudie}, \&
  {Winters}}]{Banados2018}
{Ba{\~n}ados}, E., {Venemans}, B.~P., {Mazzucchelli}, C., {et~al.} 2018, \nat,
  553, 473

\bibitem[{{Ba{\~n}ados} {et~al.}(2015){Ba{\~n}ados}, {Venemans}, {Morganson},
  {Hodge}, {Decarli}, {Walter}, {Stern}, {Schlafly}, {Farina}, {Greiner},
  {Chambers}, {Fan}, {Rix}, {Burgett}, {Draper}, {Flewelling}, {Kaiser},
  {Metcalfe}, {Morgan}, {Tonry}, \& {Wainscoat}}]{Banados2015}
{Ba{\~n}ados}, E., {Venemans}, B.~P., {Morganson}, E., {et~al.} 2015, \apj,
  804, 118

\bibitem[{{Belladitta} {et~al.}(2020){Belladitta}, {Moretti}, {Caccianiga},
  {Spingola}, {Severgnini}, {Della Ceca}, {Ghisellini}, {Dallacasa},
  {Sbarrato}, {Cicone}, {Cassar{\`a}}, \& {Pedani}}]{Belladitta2020}
{Belladitta}, S., {Moretti}, A., {Caccianiga}, A., {et~al.} 2020, \aap, 635, L7

\bibitem[{{Blandford} {et~al.}(2019){Blandford}, {Meier}, \&
  {Readhead}}]{Blandford2019}
{Blandford}, R., {Meier}, D., \& {Readhead}, A. 2019, \araa, 57, 467

\bibitem[{Chambers {et~al.}(2016)Chambers, Magnier, Metcalfe, Flewelling,
  Huber, Waters, Denneau, Draper, Farrow, Finkbeiner, Holmberg, Koppenhoefer,
  Price, Rest, Saglia, Schlafly, Smartt, Sweeney, Wainscoat, Burgett, Chastel,
  Grav, Heasley, Hodapp, Jedicke, Kaiser, Kudritzki, Luppino, Lupton, Monet,
  Morgan, Onaka, Shiao, Stubbs, Tonry, White, Ba{\~{n}}ados, Bell, Bender,
  Bernard, Boegner, Boffi, Botticella, Calamida, Casertano, Chen, Chen, Cole,
  Deacon, Frenk, Fitzsimmons, Gezari, Gibbs, Goessl, Goggia, Gourgue, Goldman,
  Grant, Grebel, Hambly, Hasinger, Heavens, Heckman, Henderson, Henning,
  Holman, Hopp, Ip, Isani, Jackson, Keyes, Koekemoer, Kotak, Le, Liska, Long,
  Lucey, Liu, Martin, Masci, McLean, Mindel, Misra, Morganson, Murphy, Obaika,
  Narayan, Nieto-Santisteban, Norberg, Peacock, Pier, Postman, Primak, Rae,
  Rai, Riess, Riffeser, Rix, R{\"{o}}ser, Russel, Rutz, Schilbach, Schultz,
  Scolnic, Strolger, Szalay, Seitz, Small, Smith, Soderblom, Taylor, Thomson,
  Taylor, Thakar, Thiel, Thilker, Unger, Urata, Valenti, Wagner, Walder,
  Walter, Watters, Werner, Wood-Vasey, \& Wyse}]{Chambers2016}
Chambers, K.~C., Magnier, E.~A., Metcalfe, N., {et~al.} 2016, eprint
  arXiv:1612.05560 [\eprint[arXiv]{1612.05560}]

\bibitem[{{Condon} {et~al.}(2002){Condon}, {Cotton}, \&
  {Broderick}}]{Condon2002}
{Condon}, J.~J., {Cotton}, W.~D., \& {Broderick}, J.~J. 2002, \aj, 124, 675

\bibitem[{Condon {et~al.}(1998)Condon, Cotton, Greisen, Yin, Perley, Taylor, \&
  Broderick}]{Condon1998}
Condon, J.~J., Cotton, W.~D., Greisen, E.~W., {et~al.} 1998, \aj, 115, 1693

\bibitem[{Coppejans {et~al.}(2017)Coppejans, van Velzen, Intema, M{\"{u}}ller,
  Frey, Coppejans, Cseh, Williams, Falcke, K{\"{o}}rding, Orr{\'{u}}, Paragi,
  \& Gab{\'{a}}nyi}]{Coppejans2017}
Coppejans, R., van Velzen, S., Intema, H.~T., {et~al.} 2017, \mnras, 467, 2039

\bibitem[{{Dark Energy Survey Collaboration} {et~al.}(2016){Dark Energy Survey
  Collaboration}, {Abbott}, {Abdalla}, {Aleksi{\'c}}, {Allam}, {Amara},
  {Bacon}, {Balbinot}, {Banerji}, {Bechtol}, {Benoit-L{\'e}vy}, {Bernstein},
  {Bertin}, {Blazek}, {Bonnett}, {Bridle}, {Brooks}, {Brunner}, {Buckley-Geer},
  {Burke}, {Caminha}, {Capozzi}, {Carlsen}, {Carnero-Rosell}, {Carollo},
  {Carrasco-Kind}, {Carretero}, {Castander}, {Clerkin}, {Collett}, {Conselice},
  {Crocce}, {Cunha}, {D'Andrea}, {da Costa}, {Davis}, {Desai}, {Diehl},
  {Dietrich}, {Dodelson}, {Doel}, {Drlica-Wagner}, {Estrada}, {Etherington},
  {Evrard}, {Fabbri}, {Finley}, {Flaugher}, {Foley}, {Fosalba}, {Frieman},
  {Garc{\'\i}a-Bellido}, {Gaztanaga}, {Gerdes}, {Giannantonio}, {Goldstein},
  {Gruen}, {Gruendl}, {Guarnieri}, {Gutierrez}, {Hartley}, {Honscheid}, {Jain},
  {James}, {Jeltema}, {Jouvel}, {Kessler}, {King}, {Kirk}, {Kron}, {Kuehn},
  {Kuropatkin}, {Lahav}, {Li}, {Lima}, {Lin}, {Maia}, {Makler}, {Manera},
  {Maraston}, {Marshall}, {Martini}, {McMahon}, {Melchior}, {Merson}, {Miller},
  {Miquel}, {Mohr}, {Morice-Atkinson}, {Naidoo}, {Neilsen}, {Nichol}, {Nord},
  {Ogando}, {Ostrovski}, {Palmese}, {Papadopoulos}, {Peiris}, {Peoples},
  {Percival}, {Plazas}, {Reed}, {Refregier}, {Romer}, {Roodman}, {Ross},
  {Rozo}, {Rykoff}, {Sadeh}, {Sako}, {S{\'a}nchez}, {Sanchez}, {Santiago},
  {Scarpine}, {Schubnell}, {Sevilla-Noarbe}, {Sheldon}, {Smith}, {Smith},
  {Soares-Santos}, {Sobreira}, {Soumagnac}, {Suchyta}, {Sullivan}, {Swanson},
  {Tarle}, {Thaler}, {Thomas}, {Thomas}, {Tucker}, {Vieira}, {Vikram},
  {Walker}, {Wechsler}, {Weller}, {Wester}, {Whiteway}, {Wilcox}, {Yanny},
  {Zhang}, \& {Zuntz}}]{Abbott2016}
{Dark Energy Survey Collaboration}, {Abbott}, T., {Abdalla}, F.~B., {et~al.}
  2016, \mnras, 460, 1270

\bibitem[{{Decarli} {et~al.}(2018){Decarli}, {Walter}, {Venemans},
  {Ba{\~n}ados}, {Bertoldi}, {Carilli}, {Fan}, {Farina}, {Mazzucchelli},
  {Riechers}, {Rix}, {Strauss}, {Wang}, \& {Yang}}]{Decarli2018}
{Decarli}, R., {Walter}, F., {Venemans}, B.~P., {et~al.} 2018, \apj, 854, 97

\bibitem[{{Driver} {et~al.}(2011){Driver}, {Hill}, {Kelvin}, {Robotham},
  {Liske}, {Norberg}, {Baldry}, {Bamford}, {Hopkins}, {Loveday}, {Peacock},
  {Andrae}, {Bland-Hawthorn}, {Brough}, {Brown}, {Cameron}, {Ching}, {Colless},
  {Conselice}, {Croom}, {Cross}, {de Propris}, {Dye}, {Drinkwater}, {Ellis},
  {Graham}, {Grootes}, {Gunawardhana}, {Jones}, {van Kampen}, {Maraston},
  {Nichol}, {Parkinson}, {Phillipps}, {Pimbblet}, {Popescu}, {Prescott},
  {Roseboom}, {Sadler}, {Sansom}, {Sharp}, {Smith}, {Taylor}, {Thomas},
  {Tuffs}, {Wijesinghe}, {Dunne}, {Frenk}, {Jarvis}, {Madore}, {Meyer},
  {Seibert}, {Staveley-Smith}, {Sutherland}, \& {Warren}}]{Driver2011}
{Driver}, S.~P., {Hill}, D.~T., {Kelvin}, L.~S., {et~al.} 2011, \mnras, 413,
  971

\bibitem[{{Drouart} {et~al.}(2020){Drouart}, {Seymour}, {Galvin}, {Afonso},
  {Callingham}, {De Breuck}, {Johnston-Hollitt}, {Kapi{\'n}ska}, {Lehnert}, \&
  {Vernet}}]{Drouart2020}
{Drouart}, G., {Seymour}, N., {Galvin}, T.~J., {et~al.} 2020, \pasa, 37, e026

\bibitem[{{Edge} {et~al.}(2013){Edge}, {Sutherland}, {Kuijken}, {Driver},
  {McMahon}, {Eales}, \& {Emerson}}]{Edge2013}
{Edge}, A., {Sutherland}, W., {Kuijken}, K., {et~al.} 2013, The Messenger, 154,
  32

\bibitem[{{Fan} {et~al.}(2019){Fan}, {Wang}, {Yang}, {Keeton}, {Yue},
  {Zabludoff}, {Bian}, {Bonaglia}, {Georgiev}, {Hennawi}, {Li}, {McGreer},
  {Naidu}, {Pacucci}, {Rabien}, {Thompson}, {Venemans}, {Walter}, {Wang}, \&
  {Wu}}]{Fan2019}
{Fan}, X., {Wang}, F., {Yang}, J., {et~al.} 2019, \apjl, 870, L11

\bibitem[{{Fomalont}(1999)}]{Fomalont1999}
{Fomalont}, E.~B. 1999, in Astronomical Society of the Pacific Conference
  Series, Vol. 180, Synthesis Imaging in Radio Astronomy II, ed. G.~B.
  {Taylor}, C.~L. {Carilli}, \& R.~A. {Perley}, 301

\bibitem[{{Gaikwad} {et~al.}(2020){Gaikwad}, {Rauch}, {Haehnelt}, {Puchwein},
  {Bolton}, {Keating}, {Kulkarni}, {Ir{\v{s}}i{\v{c}}}, {Ba{\~n}ados},
  {Becker}, {Boera}, {Zahedy}, {Chen}, {Carswell}, {Chardin}, \&
  {Rorai}}]{Gaikwad2020}
{Gaikwad}, P., {Rauch}, M., {Haehnelt}, M.~G., {et~al.} 2020, \mnras, 494, 5091

\bibitem[{{Hovatta} {et~al.}(2008){Hovatta}, {Nieppola}, {Tornikoski},
  {Valtaoja}, {Aller}, \& {Aller}}]{Hovatta2008}
{Hovatta}, T., {Nieppola}, E., {Tornikoski}, M., {et~al.} 2008, \aap, 485, 51

\bibitem[{{Intema} {et~al.}(2017){Intema}, {Jagannathan}, {Mooley}, \&
  {Frail}}]{Intema2017}
{Intema}, H.~T., {Jagannathan}, P., {Mooley}, K.~P., \& {Frail}, D.~A. 2017,
  \aap, 598, A78

\bibitem[{{Jiang} {et~al.}(2007){Jiang}, {Fan}, {Ivezi{\'c}}, {Richards},
  {Schneider}, {Strauss}, \& {Kelly}}]{Jiang2007}
{Jiang}, L., {Fan}, X., {Ivezi{\'c}}, {\v{Z}}., {et~al.} 2007, \apj, 656, 680

\bibitem[{{Johnston} {et~al.}(2008){Johnston}, {Taylor}, {Bailes}, {Bartel},
  {Baugh}, {Bietenholz}, {Blake}, {Braun}, {Brown}, {Chatterjee}, {Darling},
  {Deller}, {Dodson}, {Edwards}, {Ekers}, {Ellingsen}, {Feain}, {Gaensler},
  {Haverkorn}, {Hobbs}, {Hopkins}, {Jackson}, {James}, {Joncas}, {Kaspi},
  {Kilborn}, {Koribalski}, {Kothes}, {Landecker}, {Lenc}, {Lovell}, {Macquart},
  {Manchester}, {Matthews}, {McClure-Griffiths}, {Norris}, {Pen}, {Phillips},
  {Power}, {Protheroe}, {Sadler}, {Schmidt}, {Stairs}, {Staveley-Smith},
  {Stil}, {Tingay}, {Tzioumis}, {Walker}, {Wall}, \& {Wolleben}}]{Johnson2008}
{Johnston}, S., {Taylor}, R., {Bailes}, M., {et~al.} 2008, Experimental
  Astronomy, 22, 151

\bibitem[{{Kashikawa} {et~al.}(2006){Kashikawa}, {Shimasaku}, {Malkan}, {Doi},
  {Matsuda}, {Ouchi}, {Taniguchi}, {Ly}, {Nagao}, {Iye}, {Motohara},
  {Murayama}, {Murozono}, {Nariai}, {Ohta}, {Okamura}, {Sasaki}, {Shioya}, \&
  {Umemura}}]{Kashikawa2006}
{Kashikawa}, N., {Shimasaku}, K., {Malkan}, M.~A., {et~al.} 2006, \apj, 648, 7

\bibitem[{{Kellermann} {et~al.}(2016){Kellermann}, {Condon}, {Kimball},
  {Perley}, \& {Ivezi{\'c}}}]{Kellermann2016}
{Kellermann}, K.~I., {Condon}, J.~J., {Kimball}, A.~E., {Perley}, R.~A., \&
  {Ivezi{\'c}}, {\v{Z}}. 2016, \apj, 831, 168

\bibitem[{Kellermann {et~al.}(1989)Kellermann, Sramek, Schmidt, Shaffer, \&
  Green}]{Kellerman1989}
Kellermann, K.~I., Sramek, R., Schmidt, M., Shaffer, D.~B., \& Green, R. 1989,
  \aj, 98, 1195

\bibitem[{{Lacy} {et~al.}(2020){Lacy}, {Baum}, {Chandler}, {Chatterjee},
  {Clarke}, {Deustua}, {English}, {Farnes}, {Gaensler}, {Gugliucci},
  {Hallinan}, {Kent}, {Kimball}, {Law}, {Lazio}, {Marvil}, {Mao}, {Medlin},
  {Mooley}, {Murphy}, {Myers}, {Osten}, {Richards}, {Rosolowsky}, {Rudnick},
  {Schinzel}, {Sivakoff}, {Sjouwerman}, {Taylor}, {White}, {Wrobel},
  {Andernach}, {Beasley}, {Berger}, {Bhatnager}, {Birkinshaw}, {Bower},
  {Brandt}, {Brown}, {Burke-Spolaor}, {Butler}, {Comerford}, {Demorest}, {Fu},
  {Giacintucci}, {Golap}, {G{\"u}th}, {Hales}, {Hiriart}, {Hodge}, {Horesh},
  {Ivezi{\'c}}, {Jarvis}, {Kamble}, {Kassim}, {Liu}, {Loinard}, {Lyons},
  {Masters}, {Mezcua}, {Moellenbrock}, {Mroczkowski}, {Nyland}, {O'Dea},
  {O'Sullivan}, {Peters}, {Radford}, {Rao}, {Robnett}, {Salcido}, {Shen},
  {Sobotka}, {Witz}, {Vaccari}, {van Weeren}, {Vargas}, {Williams}, \&
  {Yoon}}]{Lacy2020}
{Lacy}, M., {Baum}, S.~A., {Chandler}, C.~J., {et~al.} 2020, \pasp, 132, 035001

\bibitem[{{Liu} {et~al.}(2021){Liu}, {Wang}, {Momjian}, {Ba{\~n}ados},
  {Zeimann}, {Willott}, {Matsuoka}, {Omont}, {Shao}, {Li}, \& {Li}}]{Liu2021}
{Liu}, Y., {Wang}, R., {Momjian}, E., {et~al.} 2021, \apj, 908, 124

\bibitem[{{Matsuoka} {et~al.}(2019){Matsuoka}, {Onoue}, {Kashikawa}, {Strauss},
  {Iwasawa}, {Lee}, {Imanishi}, {Nagao}, {Akiyama}, {Asami}, {Bosch},
  {Furusawa}, {Goto}, {Gunn}, {Harikane}, {Ikeda}, {Izumi}, {Kawaguchi},
  {Kato}, {Kikuta}, {Kohno}, {Komiyama}, {Koyama}, {Lupton}, {Minezaki},
  {Miyazaki}, {Murayama}, {Niida}, {Nishizawa}, {Noboriguchi}, {Oguri}, {Ono},
  {Ouchi}, {Price}, {Sameshima}, {Schulze}, {Shirakata}, {Silverman},
  {Sugiyama}, {Tait}, {Takada}, {Takata}, {Tanaka}, {Tang}, {Toba}, {Utsumi},
  {Wang}, \& {Yamashita}}]{Matsuoka2019a}
{Matsuoka}, Y., {Onoue}, M., {Kashikawa}, N., {et~al.} 2019, \apjl, 872, L2

\bibitem[{{Mauch} {et~al.}(2003){Mauch}, {Murphy}, {Buttery}, {Curran},
  {Hunstead}, {Piestrzynski}, {Robertson}, \& {Sadler}}]{Mauch2003}
{Mauch}, T., {Murphy}, T., {Buttery}, H.~J., {et~al.} 2003, \mnras, 342, 1117

\bibitem[{{Mazzucchelli} {et~al.}(2017){Mazzucchelli}, {Ba{\~n}ados},
  {Venemans}, {Decarli}, {Farina}, {Walter}, {Eilers}, {Rix}, {Simcoe},
  {Stern}, {Fan}, {Schlafly}, {De Rosa}, {Hennawi}, {Chambers}, {Greiner},
  {Burgett}, {Draper}, {Kaiser}, {Kudritzki}, {Magnier}, {Metcalfe}, {Waters},
  \& {Wainscoat}}]{Mazzucchelli2017}
{Mazzucchelli}, C., {Ba{\~n}ados}, E., {Venemans}, B.~P., {et~al.} 2017, \apj,
  849, 91

\bibitem[{{McConnell} {et~al.}(2020){McConnell}, {Hale}, {Lenc}, {Banfield},
  {Heald}, {Hotan}, {Leung}, {Moss}, {Murphy}, {O'Brien}, {Pritchard}, {Raja},
  {Sadler}, {Stewart}, {Thomson}, {Whiting}, {Allison}, {Amy}, {Anderson},
  {Ball}, {Bannister}, {Bell}, {Bock}, {Bolton}, {Bunton}, {Chippendale},
  {Collier}, {Cooray}, {Cornwell}, {Diamond}, {Edwards}, {Gupta}, {Hayman},
  {Heywood}, {Jackson}, {Koribalski}, {Lee-Waddell}, {McClure-Griffiths}, {Ng},
  {Norris}, {Phillips}, {Reynolds}, {Roxby}, {Schinckel}, {Shields},
  {Tremblay}, {Tzioumis}, {Voronkov}, \& {Westmeier}}]{McConnell2020}
{McConnell}, D., {Hale}, C.~L., {Lenc}, E., {et~al.} 2020, \pasa, 37, e048

\bibitem[{{McGreer} {et~al.}(2006){McGreer}, {Becker}, {Helfand}, \&
  {White}}]{McGreer2006}
{McGreer}, I.~D., {Becker}, R.~H., {Helfand}, D.~J., \& {White}, R.~L. 2006,
  \apj, 652, 157

\bibitem[{{McMullin} {et~al.}(2007){McMullin}, {Waters}, {Schiebel}, {Young},
  \& {Golap}}]{Mcmullin2007}
{McMullin}, J.~P., {Waters}, B., {Schiebel}, D., {Young}, W., \& {Golap}, K.
  2007, in Astronomical Society of the Pacific Conference Series, Vol. 376,
  Astronomical Data Analysis Software and Systems XVI, ed. R.~A. {Shaw},
  F.~{Hill}, \& D.~J. {Bell}, 127

\bibitem[{{Norris} {et~al.}(2011){Norris}, {Hopkins}, {Afonso}, {Brown},
  {Condon}, {Dunne}, {Feain}, {Hollow}, {Jarvis}, {Johnston-Hollitt}, {Lenc},
  {Middelberg}, {Padovani}, {Prandoni}, {Rudnick}, {Seymour}, {Umana},
  {Andernach}, {Alexander}, {Appleton}, {Bacon}, {Banfield}, {Becker}, {Brown},
  {Ciliegi}, {Jackson}, {Eales}, {Edge}, {Gaensler}, {Giovannini}, {Hales},
  {Hancock}, {Huynh}, {Ibar}, {Ivison}, {Kennicutt}, {Kimball}, {Koekemoer},
  {Koribalski}, {L{\'o}pez-S{\'a}nchez}, {Mao}, {Murphy}, {Messias},
  {Pimbblet}, {Raccanelli}, {Randall}, {Reiprich}, {Roseboom},
  {R{\"o}ttgering}, {Saikia}, {Sharp}, {Slee}, {Smail}, {Thompson}, {Urquhart},
  {Wall}, \& {Zhao}}]{Norris2011}
{Norris}, R.~P., {Hopkins}, A.~M., {Afonso}, J., {et~al.} 2011, \pasa, 28, 215

\bibitem[{{Retana-Montenegro} \& {R{\"o}ttgering}(2017)}]{Retana2017}
{Retana-Montenegro}, E. \& {R{\"o}ttgering}, H.~J.~A. 2017, \aap, 600, A97

\bibitem[{{Runnoe} {et~al.}(2012){Runnoe}, {Brotherton}, \&
  {Shang}}]{Runnoe2012}
{Runnoe}, J.~C., {Brotherton}, M.~S., \& {Shang}, Z. 2012, \mnras, 422, 478

\bibitem[{{Seymour} {et~al.}(2020){Seymour}, {Huynh}, {Shabala}, {Rogers},
  {Davies}, {Turner}, {O'Brien}, {Ishwara-Chandra}, {Thorne}, {Galvin},
  {Jarrett}, {Andernach}, {Anderson}, {Bunton}, {Chow}, {Collier}, {Driver},
  {Filipovic}, {G{\"u}rkan}, {Hopkins}, {Kapi{\'n}ska}, {Leahy}, {Marvil},
  {Manojlovic}, {Norris}, {Phillips}, {Robotham}, {Rudnick}, {Singh}, \&
  {White}}]{Seymour2020}
{Seymour}, N., {Huynh}, M., {Shabala}, S.~S., {et~al.} 2020, \pasa, 37, e013

\bibitem[{{Shen} {et~al.}(2008){Shen}, {Greene}, {Strauss}, {Richards}, \&
  {Schneider}}]{Shen2008}
{Shen}, Y., {Greene}, J.~E., {Strauss}, M.~A., {Richards}, G.~T., \&
  {Schneider}, D.~P. 2008, \apj, 680, 169

\bibitem[{{Stern} {et~al.}(2000){Stern}, {Djorgovski}, {Perley}, {de Carvalho},
  \& {Wall}}]{Stern2000}
{Stern}, D., {Djorgovski}, S.~G., {Perley}, R.~A., {de Carvalho}, R.~R., \&
  {Wall}, J.~V. 2000, \aj, 119, 1526

\bibitem[{{Vanden Berk} {et~al.}(2001){Vanden Berk}, {Richards}, {Bauer},
  {Strauss}, {Schneider}, {Heckman}, {York}, {Hall}, {Fan}, {Knapp},
  {Anderson}, {Annis}, {Bahcall}, {Bernardi}, {Briggs}, {Brinkmann}, {Brunner},
  {Burles}, {Carey}, {Castander}, {Connolly}, {Crocker}, {Csabai}, {Doi},
  {Finkbeiner}, {Friedman}, {Frieman}, {Fukugita}, {Gunn}, {Hennessy},
  {Ivezi{\'c}}, {Kent}, {Kunszt}, {Lamb}, {Leger}, {Long}, {Loveday}, {Lupton},
  {Meiksin}, {Merelli}, {Munn}, {Newberg}, {Newcomb}, {Nichol}, {Owen}, {Pier},
  {Pope}, {Rockosi}, {Schlegel}, {Siegmund}, {Smee}, {Snir}, {Stoughton},
  {Stubbs}, {SubbaRao}, {Szalay}, {Szokoly}, {Tremonti}, {Uomoto}, {Waddell},
  {Yanny}, \& {Zheng}}]{Vandenberk2001}
{Vanden Berk}, D.~E., {Richards}, G.~T., {Bauer}, A., {et~al.} 2001, \aj, 122,
  549

\bibitem[{{Venemans} {et~al.}(2018){Venemans}, {Decarli}, {Walter},
  {Ba{\~n}ados}, {Bertoldi}, {Fan}, {Farina}, {Mazzucchelli}, {Riechers},
  {Rix}, {Wang}, \& {Yang}}]{Venemans2018}
{Venemans}, B.~P., {Decarli}, R., {Walter}, F., {et~al.} 2018, \apj, 866, 159

\bibitem[{{Venemans} {et~al.}(2020){Venemans}, {Walter}, {Neeleman}, {Novak},
  {Otter}, {Decarli}, {Ba{\~n}ados}, {Drake}, {Farina}, {Kaasinen},
  {Mazzucchelli}, {Carilli}, {Fan}, {Rix}, \& {Wang}}]{Venemans2020}
{Venemans}, B.~P., {Walter}, F., {Neeleman}, M., {et~al.} 2020, \apj, 904, 130

\bibitem[{{Volonteri} {et~al.}(2015){Volonteri}, {Silk}, \&
  {Dubus}}]{Volonteri2015}
{Volonteri}, M., {Silk}, J., \& {Dubus}, G. 2015, \apj, 804, 148

\bibitem[{{Wang} {et~al.}(2020){Wang}, {Fan}, {Yang}, {Mazzucchelli}, {Wu},
  {Li}, {Banados}, {Farina}, {Nanni}, {Ai}, {Bian}, {Davies}, {Decarli},
  {Hennawi}, {Schindler}, {Venemans}, \& {Walter}}]{Wang2020}
{Wang}, F., {Fan}, X., {Yang}, J., {et~al.} 2020, arXiv e-prints,
  arXiv:2011.12458

\bibitem[{{Wang} {et~al.}(2021){Wang}, {Yang}, {Fan}, {Hennawi}, {Barth},
  {Banados}, {Bian}, {Boutsia}, {Connor}, {Davies}, {Decarli}, {Eilers},
  {Farina}, {Green}, {Jiang}, {Li}, {Mazzucchelli}, {Nanni}, {Schindler},
  {Venemans}, {Walter}, {Wu}, \& {Yue}}]{Wang2021}
{Wang}, F., {Yang}, J., {Fan}, X., {et~al.} 2021, arXiv e-prints,
  arXiv:2101.03179

\bibitem[{{Wang} {et~al.}(2019){Wang}, {Yang}, {Fan}, {Wu}, {Yue}, {Li},
  {Bian}, {Jiang}, {Ba{\~n}ados}, {Schindler}, {Findlay}, {Davies}, {Decarli},
  {Farina}, {Green}, {Hennawi}, {Huang}, {Mazzuccheli}, {McGreer}, {Venemans},
  {Walter}, {Dye}, {Lyke}, {Myers}, \& {Haze Nunez}}]{Wang2019}
{Wang}, F., {Yang}, J., {Fan}, X., {et~al.} 2019, \apj, 884, 30

\bibitem[{{Wenger} {et~al.}(2000){Wenger}, {Ochsenbein}, {Egret}, {Dubois},
  {Bonnarel}, {Borde}, {Genova}, {Jasniewicz}, {Lalo{\"e}}, {Lesteven}, \&
  {Monier}}]{Simbad2000}
{Wenger}, M., {Ochsenbein}, F., {Egret}, D., {et~al.} 2000, \aaps, 143, 9

\bibitem[{{Willott} {et~al.}(2010){Willott}, {Delorme}, {Reyl{\'e}}, {Albert},
  {Bergeron}, {Crampton}, {Delfosse}, {Forveille}, {Hutchings}, {McLure},
  {Omont}, \& {Schade}}]{Willot2010}
{Willott}, C.~J., {Delorme}, P., {Reyl{\'e}}, C., {et~al.} 2010, \aj, 139, 906

\bibitem[{{Yang} {et~al.}(2020{\natexlab{a}}){Yang}, {Wang}, {Fan}, {Hennawi},
  {Davies}, {Yue}, {Banados}, {Wu}, {Venemans}, {Barth}, {Bian}, {Boutsia},
  {Decarli}, {Farina}, {Green}, {Jiang}, {Li}, {Mazzucchelli}, \&
  {Walter}}]{Yang2020}
{Yang}, J., {Wang}, F., {Fan}, X., {et~al.} 2020{\natexlab{a}}, \apjl, 897, L14

\bibitem[{{Yang} {et~al.}(2020{\natexlab{b}}){Yang}, {Wang}, {Fan}, {Hennawi},
  {Davies}, {Yue}, {Eilers}, {Farina}, {Wu}, {Bian}, {Pacucci}, \&
  {Lee}}]{Yang2020b}
{Yang}, J., {Wang}, F., {Fan}, X., {et~al.} 2020{\natexlab{b}}, \apj, 904, 26

\bibitem[{{Zhu} {et~al.}(2020){Zhu}, {Brandt}, {Luo}, {Wu}, {Xue}, \&
  {Yang}}]{Zhu2020}
{Zhu}, S.~F., {Brandt}, W.~N., {Luo}, B., {et~al.} 2020, \mnras, 496, 245

\end{thebibliography}
%

\end{document}